# Multi-contrast wide-field mid-infrared photothermal imaging

ANOOJ THAYYIL RAVEENDRAN,[1] CORNELIA REUTER,[1,2,3] SAMIR F. EL-MASHTOLY,[1] JÜRGEN POPP,[1,2] AND CHRISTOPH KRAFFT[1,*]

[1]*Leibniz Institute of Photonic Technology, Member of Leibniz Research Alliance Leibniz Health Technologies, Member of the Leibniz Center for Photonics in Infection Research, Albert-Einstein-Str. 9, 07745 Jena, Germany*
[2]*Friedrich-Schiller-University Jena, Institute of Physical Chemistry, Member of Leibniz Research Alliance Leibniz Health Technologies, Member of the Leibniz Center for Photonics in Infection Research, Helmholtzweg 4, 07743 Jena, Germany*
[3]*InfectoGnostics Research Campus Jena, Center for Applied Research, Philosophenweg 7, 07743 Jena, Germany*

**Corresponding author: Christoph.Krafft@leibniz-ipht.de*

**Abstract:** Wide-field mid-infrared photothermal (MIP) imaging enables high-speed, chemically specific imaging at sub-micrometer spatial resolution through several possible contrast mechanisms, including scattering change, fluorescence modulation, and phase change. Each mechanism offers distinct advantages, yet existing wide-field MIP instruments have typically been limited to a single contrast mechanism so far, requiring a dedicated instrument for each. Here, we introduce a switchable multi-contrast wide-field MIP microscope that allows the photothermal readout mechanism (based on scattering-, fluorescence-, or quantitative phase imaging (QPI)) to be matched to the sample and imaging conditions. We characterize the spectral fidelity and signal-to-noise (SNR) ratio of each mechanism using 5 µm polystyrene (PS) beads, and the effective spatial resolution using 300 nm PS beads. Scattering- and fluorescence-based detection achieved a sub-500 nm measured point spread function full width at half maximum, while QPI-based detection reached approximately 900 nm but provided up to 26-fold higher SNR than scattering. We further demonstrate imaging across a range of samples, including A549 cells in buffer, mouse lung tissue section, bacteria, and SU8 polymer at 1.4 and 5 µm thickness, with a field of view exceeding 200 µm. Comparison across these samples shows that each contrast mechanism offers distinct trade-offs in resolution, sensitivity, and applicability, guiding readout-strategy selection by sample type. These results establish a flexible, switchable multi-contrast platform for wide-field MIP imaging, with broad applicability across polymer and biological samples.

## 1. Introduction

Mid-infrared photothermal (MIP) imaging has emerged as a label-free chemical imaging technique that combines the molecular specificity of mid-infrared (IR) absorption with the sub-micron spatial resolution of visible-light microscopy. In its wide-field implementation, MIP imaging illuminates and detects a field of view (FOV) typically exceeding 30 µm, enabling fast acquisition and large-area sample characterization [1, 2]. The photothermal signal generated by mid-IR absorption can be read out through several distinct optical contrast mechanisms, including scattering-based approaches such as dark-field imaging, phase-sensitive imaging, fluorescence modulation, photoacoustic detection, and other interferometric schemes [3-7]. Each of these mechanisms offers a particular advantage: for instance, scattering-based detection allows for a simple optical layout, fluorescence modulation provides high modulation depth, and quantitative phase imaging (QPI) enables imaging under low mid-IR fluence, with optimal choice depending on the sample properties and imaging conditions. Despite this diversity of contrast mechanisms, existing wide-field MIP systems are generally built around a single detection modality.

Prior work demonstrated wide-field MIP imaging using scattering-based detection, phase-based detection, and fluorescence modulation, each in a separate dedicated system. Scattering-based configurations have been used to achieve simple, lens-based photothermal detection without the need for an interferometric reference arm, making these systems straightforward to implement and align. This approach was applied across a range of samples, including sub-micrometer particles, single virus imaging, bacterial imaging, and tissue screening [8-12]. These studies refined the sensitivity and practicality of the approach, employing nanosecond visible pulses for precise temporal gating of the photothermal signal and high-full-well cameras combined with frame averaging to improve signal-to-noise ratio (SNR). Beyond scattering, phase-sensitive detection has been explored for wide-field MIP, since it captures the photothermal effect as a change in optical path length rather than scattered intensity. This makes QPI suitable for optically transparent, weakly scattering samples such as live cells, where it additionally recovers dry-mass and morphological information from the phase image itself. QPI-based methods facilitated video-rate live-cell imaging and were extended to nanoscale resolution to dissect bacterial populations [13, 14]. More recently, it was extended to hyperspectral imaging of live cells over large FOV and at high confluency, using low illumination power to minimize photodamage while imaging larger cell populations [15]. Fluorescence modulation offers another route to wide-field MIP contrast, where mid-IR-induced heating changes a fluorophore's quantum yield rather than its scattering or phase [6, 7]. This thermal quenching effect is roughly two orders of magnitude stronger than the corresponding change in scattering intensity, giving fluorescence detection higher sensitivity while avoiding the speckle and coherence artifacts inherent to scattering-based schemes [16]. Fluorescence-detected mid-IR photothermal (F-MIP) imaging facilitated volumetric imaging of protein content in bacteria and lipid droplets in living cancer cells, including metabolic changes linked to drug resistance [17], and, more recently, millimeter-scale FOV imaging of tuberculosis-infected tissue and microplastic particles using high-pulse-energy sources [18]. However, each implementation remains restricted to a single contrast mechanism, limiting the ability to match the imaging modality to the sample and application at hand.

Here, we address this limitation by developing a wide-field MIP platform that integrates scattering, QPI, and fluorescence-modulation contrast mechanisms within a single optical setup. This switchable design allows the readout mechanism to be selected according to the sample and imaging task, rather than requiring a dedicated instrument for each contrast mechanism, while also enabling direct comparison of contrast mechanisms on the same sample and FOV. We characterize the effective spatial resolution, SNR, FOV, and imaging speed of each contrast mechanism using PS beads (300 nm for spatial resolution, 5 μm for SNR and spectral fidelity), and demonstrate the platform's versatility across a diverse set of samples, including A549 cells, bacteria, structured polymer thin films, and tissue sections. We further provide a qualitative and quantitative comparison of the three contrast mechanisms, highlighting their relative strengths and trade-offs across the samples studied.

## 2. Materials and methods

### *2.1 Instrumentation and optics for wide-field MIP measurements*

The instrumental setup of the multi-contrast wide-field MIP microscope is shown in Fig. 1. The mid-IR excitation was provided by a picosecond optical parametric oscillator (OPO, PT501, Ekspla), operating at a 100 Hz repetition rate with a pulse width of ~26 ps (pulse energy up to 360 μJ) and bandwidth <4 $cm^{-1}$, which additionally provides a fixed 1064 nm output with a similar pulse width. The mid-IR beam was guided through a series of gold mirrors, modulated by an optical chopper/internal gating (Thorlabs MC1F2), and focused onto the sample using an off-axis parabolic mirror (OAP1) with a focal length (FL) of either 5 cm, 7.5 cm, or 15 cm, depending on the desired FOV.

The visible imaging path incorporates two light sources, a 450 nm LED (UHP-T-450-SR) or a 532 nm laser, selected depending on the contrast mechanism used. For fluorescence imaging, the LED is used rather than the laser, as its lower pulse energy and longer pulse width (<5 µs) reduce photobleaching. The LED is guided via mirror M1, passing through excitation filter F1 (bandpass, 450 ± 20 nm) to define the excitation band, into microscope objective MO1 (Zeiss 50×, 0.75 NA or Zeiss 20×, 0.45 NA water immersion), while the path toward objective MO2 (Zeiss 20×, 0.4 NA) is blocked. MO1 collects the resulting fluorescence and focuses it onto an emission filter F2 (longpass, 500 nm) to reject residual excitation light before imaging it onto a CMOS camera (Ximea, 24 MP, MX-245MG-SY-X4G3-FF). Depending on the desired collection efficiency, the fluorescence signal can be directed to the camera using either a beam splitter (BS, Edmund Optics, 35 mm × 35 mm, 50R/50T) or a dichroic mirror (DM, Thorlabs, MD480). For scattering and phase-based imaging, a visible laser source of 532 nm is used instead of the LED. This beam is generated by frequency-doubling the 1064 nm OPO output to 532 nm using an LBO (EKSMA, LBO-301) crystal and is then coupled through a multimode fiber (MMF, Thorlabs M15L20) to make it partially coherent, reducing speckle noise. In this configuration, M1 is flipped out of the beam path to direct the 532 nm beam into MO1. For scattering-based detection, the visible beam passes through MO1, is reflected from the sample back toward the camera, with filter F2 removed from the detection path. For QPI, the path toward MO2 is instead opened, forming a Linnik interferometer, with a tilted reference mirror (M10), mounted on a linear stage (LS), introducing a spatial carrier frequency for off-axis phase retrieval.

The time-gated detection scheme used to extract the photothermal signal has been described in detail in our previous publication [18]. The mid-IR beam can be modulated either mechanically, using the optical chopper, or through the internal gating of the mid-IR laser itself; the chopper is limited to modulation frequencies above 5 Hz.

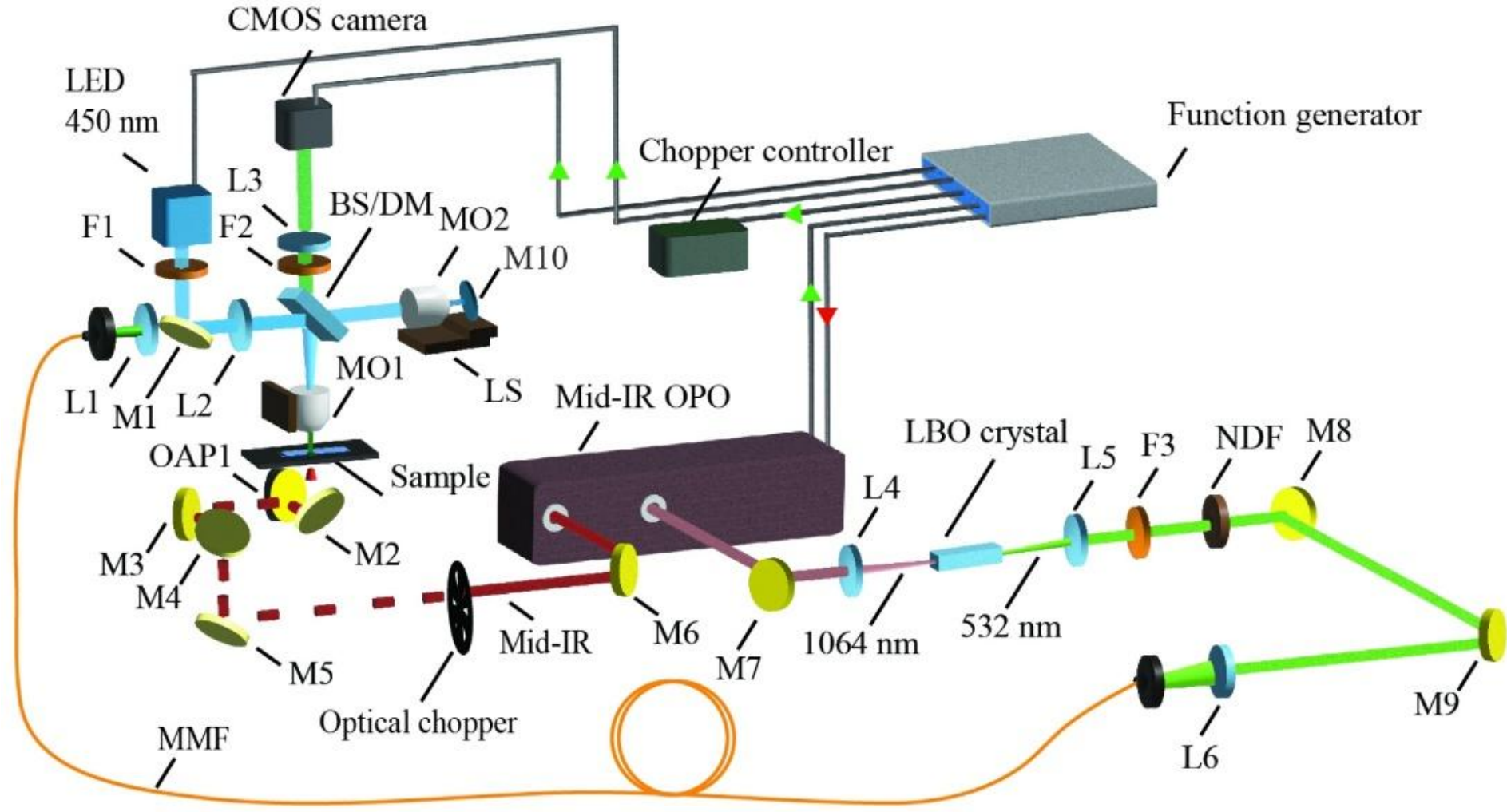

**Fig. 1.** Optical and electronic setup of the wide-field mid-IR instrument. The mid-IR beam is delivered from below and the visible light from above, in a counter-propagating configuration. Two visible sources are used: a 450 nm LED and a 532 nm source. The 532 nm beam is frequency-doubled from the OPO pump laser (1064 nm, 26 ps pulse width, 4 cm$^{-1}$ bandwidth); L4 focuses the beam into the LBO crystal for second-harmonic generation, and L5 collimates the resulting output. A bandpass filter (F3) removes residual 1064 nm beam, and L6 couples the 532 nm beam into a multimode fiber (MMF) of 20 m length to reach the detection path, where it is focused onto the back focal plane of MO1. MO1 enables both scattering and fluorescence modulation mechanisms; MO2 is used exclusively for the QPI path, forming part of the Linnik interferometer. MO1 also collects light from the sample and focuses it onto the CMOS camera via lens L3. The 450 nm LED is used only for fluorescence mode. The mid-IR beam is guided by a series of gold mirrors (M2-6) and focused onto the sample using an OAP1.

*2.2 Sample preparation*

PS beads of two different diameters were used as calibration standards for photothermal contrast: 5 µm fluorescent beads (FluoroMax Green, Thermo Fisher Scientific) and 300 nm fluorescent beads (epruibiotech, green fluorescent). Bead suspensions were diluted in 30 mL of deionized water and deposited onto a Si substrate, chosen for its reflectivity match to the Si reference mirror M10 used in the QPI channel, before being air-dried at room temperature to immobilize individual particles for imaging.

A549 human lung adenocarcinoma cells, purchased from the Deutsche Sammlung von Mikroorganismen und Zellkulturen (DSMZ, (ACC 107)), were cultured under standard conditions in Dulbecco's Modified Eagle Medium (DMEM) with high glucose and GlutaMAX™ (Thermo Fisher Scientific/Gibco, Cat# 61965-500), supplemented with 10% fetal bovine serum (Biochrom Superior) and 1% Penicillin-Streptomycin (Thermo Fisher Scientific/Gibco, Cat# 15140-122). Cells were maintained in T75 flasks at 37 °C in a humidified incubator with 5% $CO_2$ and passaged every 2–3 days at a 1:10 ratio upon reaching ~80% confluence (approximately $5 \times 10^6$ cells per flask), using washing with PBS followed by detachment with 0.05% Trypsin-EDTA (Thermo Fisher Scientific/Gibco, Cat# 25300-025) for 3 min at 37 °C. For fluorescence staining, cells grown to 60–80% confluency were incubated with SYTO 9 green-fluorescent nucleic acid stain at a final concentration of 5 µM for 30 min at 37 °C, 5% $CO_2$, protected from light. Following staining, cells were washed once with PBS, detached with 0.05% Trypsin-EDTA (3 min, 37 °C), collected by centrifugation (125 × g, 5 min), fixed with 4% paraformaldehyde in PBS for 15 min at room temperature, and washed three times with PBS prior to imaging.

5-week-old female BALB/cJRj-wild-type mice (Janvier, Le Genest-Saint-Isle, France, local ethics committee of the Thuringian State Office for Consumer Protection (TV-Nr: 02-018/16) were housed according to the institutional guidelines. The mice were sacrificed by an overdose of ketamine and xylazine. The lungs were extracted, embedded in optimum cutting temperature medium, and frozen. A 16 µm tissue section was cut with a cryotome, transferred onto a 1 mm thick $CaF_2$ window, and dried before IR measurement.

An SU8 polymer resolution target with feature thicknesses of 1.4 µm and 5 µm was fabricated in-house in the cleanroom facility at the Leibniz Institute of Photonic Technology (IPHT), Jena, using standard photolithographic patterning of SU8 photoresist on 200 µm double-side polished 10×10 mm silicon windows.

For bacterial samples, *Escherichia coli (E. coli) DSM 423* and *Staphylococcus aureus (S. aureus)* DSM 20231 were grown from overnight cultures diluted 1:20 into fresh liquid medium and incubated at 37 °C for 1 h until reaching an $OD_{600}$ of 0.5–0.7. Cells were fluorescently labeled using the *LIVE/DEAD® BacLight™ Bacterial Viability Kit* (Invitrogen™, Thermo Fisher Scientific, Waltham, MA, USA) by adding 2 µL of Component A (Syto9) to 2 mL of the main culture and incubating at 37 °C for 25–30 min. The suspension was then centrifuged, the supernatant discarded, and the pellet resuspended in 1 mL deionized water before adjusting to a final $OD_{600}$ of 1.0. Colony-forming units per milliliter (CFU/mL) were determined by plating dilutions (1:100,000) on CHROMagar Orientation™ agar plates (MastGroup GmbH, Reinfeld, Germany). Finally, 1 µL aliquots ($10^9$ and $10^7$ CFU/mL) were deposited onto precleaned $CaF_2$ (200 µm thick) slides and dried in the dark.

*2.3 Software*

Image acquisition and instrument control were performed using LabVIEW 2024 (National Instruments). For QPI, phase information was extracted from the acquired interferograms using the Fourier fringe analysis method, involving Fourier transformation, isolation of the first-order sideband, and inverse transformation to recover the phase [19]. ImageJ was used for further processing and display.

## 3. Results

### *3.1 Imaging of 5 μm PS beads and spectral fidelity*

To assess the spectral fidelity of the scattering, fluorescence, and phase contrast mechanisms, 5 μm PS beads were imaged across a range of wavenumbers spanning the C–H absorption band (Fig. 2). Hyperspectral wide-field MIP images of single beads were acquired from 1410 to 1510 $cm^{-1}$ in 10 $cm^{-1}$ steps (5 cm FL OAP1 in Fig. 1). Scattering and phase images were acquired at 20 fps, while fluorescence images (using the same BS as in Fig. 1) were acquired at 5 fps, averaging over 2.5 s to improve SNR.

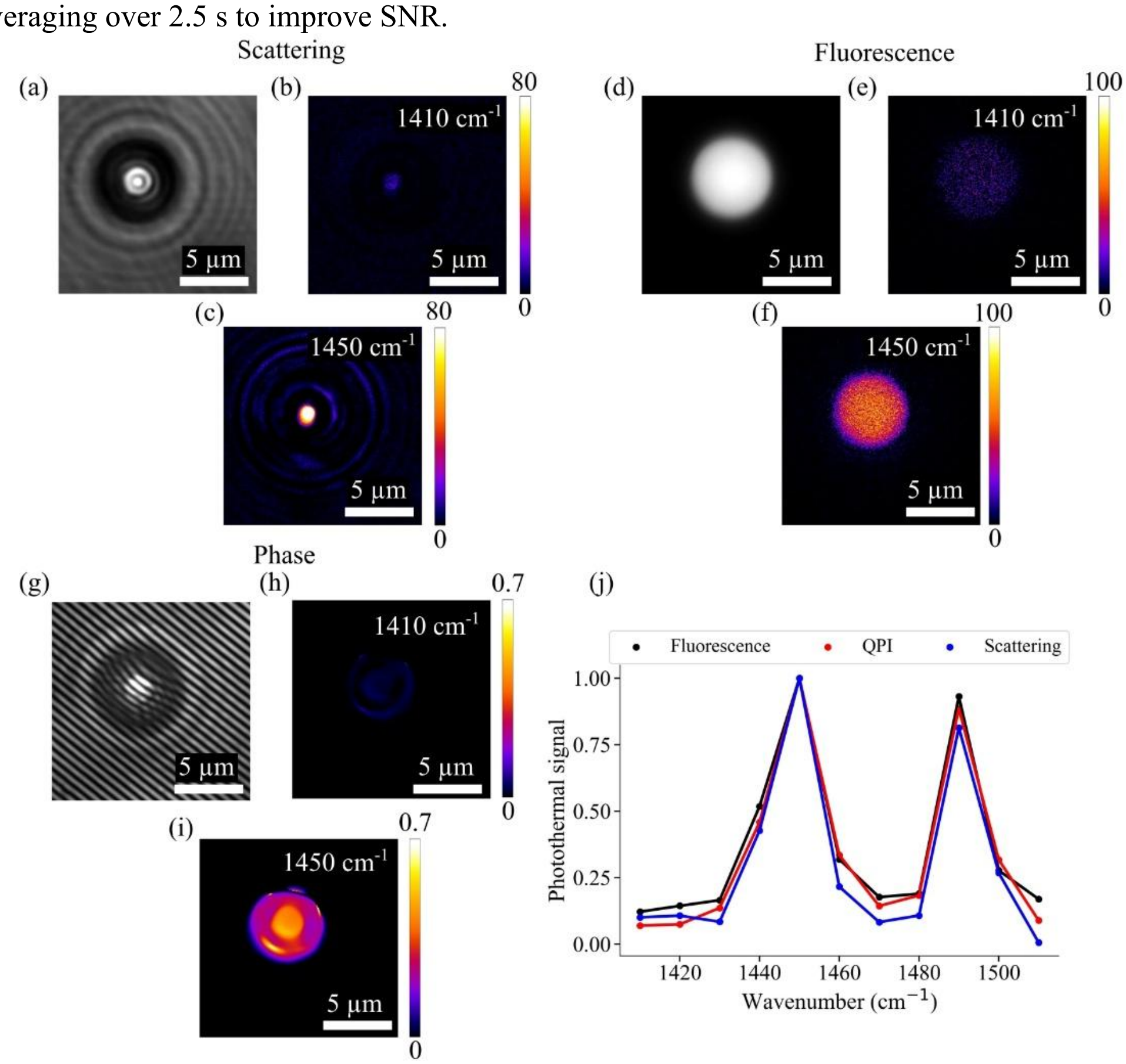

**Fig. 2.** Spectral fidelity of scattering, fluorescence, and phase contrast mechanisms in 5 μm PS beads. (a) Brightfield image of a PS bead. (b, c) Scattering-based photothermal images at 1410 $cm^{-1}$ and 1450 $cm^{-1}$, respectively. (d) Fluorescence image, with corresponding F-MIP images at 1410 $cm^{-1}$ (e) and 1450 $cm^{-1}$ (f). (g) Interference image, with corresponding QPI-based contrast images at 1410 $cm^{-1}$ (h) and 1450 $cm^{-1}$ (i). (j) Spectral response from 1410 to 1510 $cm^{-1}$ measured by the three contrast mechanisms.

Representative images at 1410 $cm^{-1}$ and 1450 $cm^{-1}$ are shown for scattering (Fig. 2a–c), fluorescence (Fig. 2d–f), and phase (Fig. 2g–i), alongside the corresponding visible images. All three contrast mechanisms show strong photothermal contrast at 1450 $cm^{-1}$ and weak contrast at 1410 $cm^{-1}$, consistent with the expected PS absorption spectrum [20]. SNR was estimated as the signal from a 5 μm circular region on the bead at 1450 $cm^{-1}$ divided by the standard deviation of the background from an equivalent area, yielding values of 5, 22, and 128 for scattering, fluorescence, and QPI, respectively. At 1450 $cm^{-1}$, scattering-based imaging shows a bright center with weak contrast at the bead edge, whereas fluorescence- and QPI-based imaging display a brighter center with contrast decreasing progressively toward the edge.

Moreover, the spectral response across this range was also calculated for each contrast mechanism and is shown in Fig. 2j, where all three show a similar trend. A minor variation is observed around 1490 $cm^{-1}$, which may be attributed to water vapor absorption of IR in this wavenumber range [21].

### *3.2 Spatial resolution*

Spatial resolution was estimated using 300 nm green-fluorescent PS beads. For this measurement, the setup used a 5 cm focal length OAP1 (Fig. 1) and a dichroic mirror (DM) in place of the beam splitter (BS). Imaging was performed with a 0.75 NA objective at 532 nm for scattering and phase contrast, and at the corresponding green fluorescence emission wavelength for the fluorescence channel. Based on the Abbe diffraction limit ($d = \lambda/2NA$), the theoretical resolution at 532 nm and 0.75 NA is approximately 355 nm; since the 300 nm bead diameter is comparable to, but still below, this diffraction limit, it provides a practical near-point-source target for experimental point spread function (PSF) characterization, though the finite bead size is expected to contribute modestly to the measured FWHM. Beads were imaged at 1450 $cm^{-1}$ using all three contrast mechanisms (Fig. 3a–c), with images averaged over 5 s. Line profiles were taken across individual beads in the 1450 $cm^{-1}$ images for each mechanism (Fig. 3d–f) and fitted with a Gaussian function to extract the full width at half maximum (FWHM). The resulting measured PSF FWHM values were 376 nm for scattering, 429 nm for fluorescence, and 904 nm for QPI, representing experimental resolution estimates for each contrast mechanism.

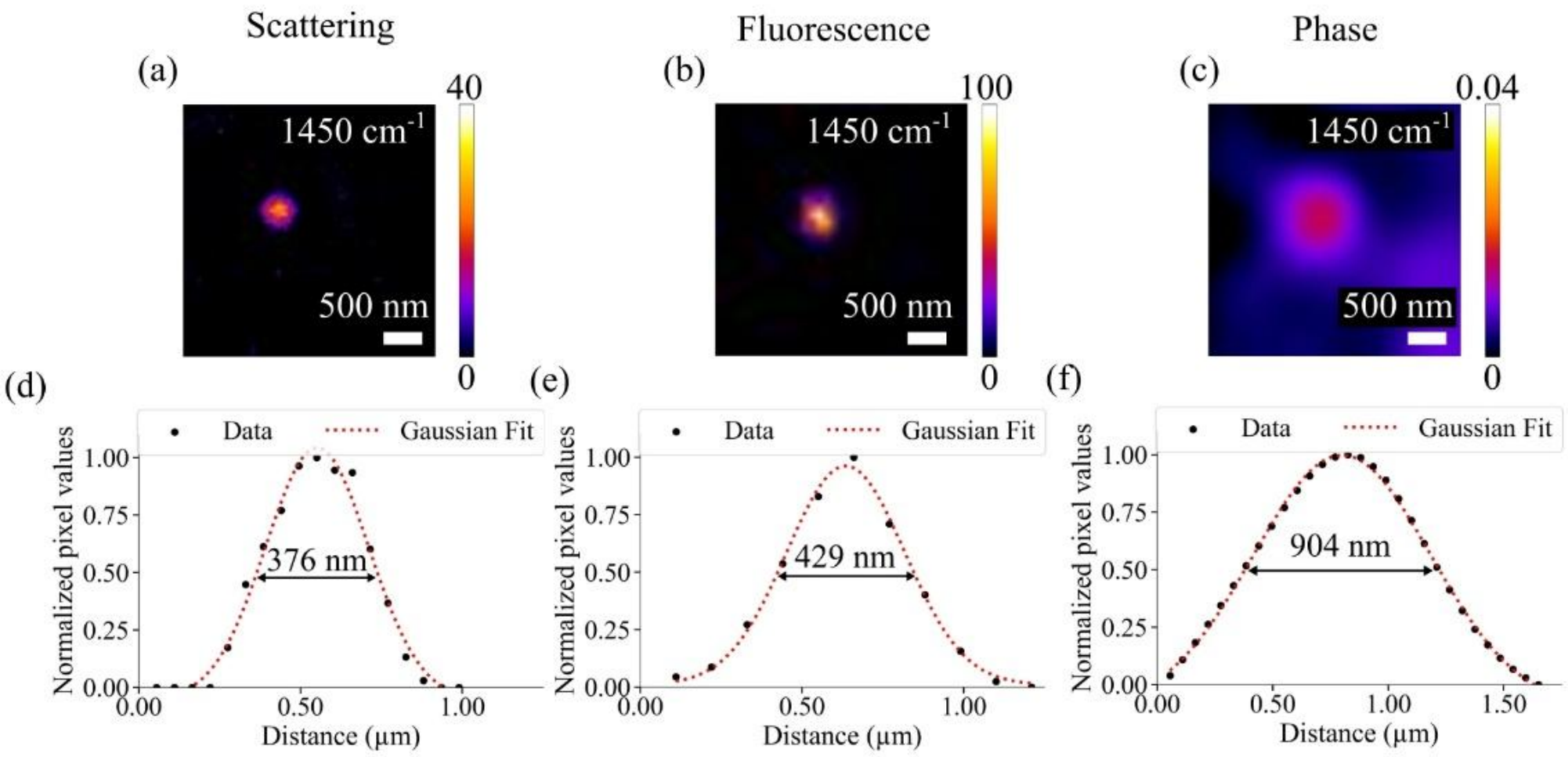

**Fig. 3.** Spatial resolution characterization using 300 nm PS beads at 1450 $cm^{-1}$. (a-c) Photothermal images of individual beads for scattering, fluorescence (at 2 fps), and QPI, respectively. (d-f) Corresponding line profiles with Gaussian fits used to extract the FWHM.

### *3.3 Biological cell imaging in buffer*

For cellular imaging, fixed A549 cells (Fig. 4a) suspended in buffer were deposited onto a silicon substrate and immobilized by sealing with a 200 μm-thick $CaF_2$ window placed on top. Cell nuclei were labeled with the SYTO 9 nucleic acid stain as described above. Wide-field MIP images were acquired with the mid-IR pump tuned to 1240 $cm^{-1}$ (Fig. 4b–d), the asymmetric $PO_2^-$ stretching band, which arises from the phosphate backbone of nucleic acids and thus provides strong contrast from the nucleus, but is also present in the phosphate head groups of membrane phospholipids; the band is therefore not exclusively nuclear. The probe beam was focused onto the sample using a 5 cm FL OAP1 as the focusing optic, and images were collected through a 20×, 0.45 NA water-immersion objective. For each detection mechanism (scattering, fluorescence, and QPI), hot and cold frames were acquired and averaged over an acquisition window of approximately 10 s to improve SNR.

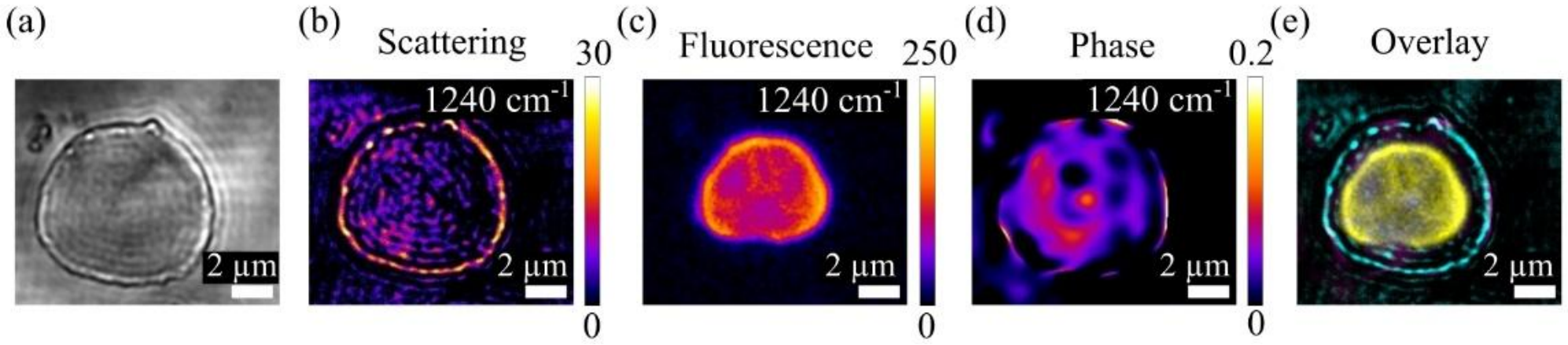

**Fig. 4.** Wide-field MIP images of an A549 cell in aqueous buffer at 1240 cm⁻¹ acquired with (a) optical brightfield contrast, (b) scattering-based detection, (c) fluorescence (2 fps, using DM in Fig. 1), and (d) QPI-based detection. (e) Composite overlay of the three MIP contrast channels, with fluorescence shown in yellow, QPI in magenta, and scattering in cyan, highlighting the spatial correlation between the signals across detection modalities.

Imaging cells in an aqueous buffer environment introduces an additional background photothermal signal from the surrounding water itself, which is particularly pronounced in the QPI channel due to its higher sensitivity to optical path length changes accumulated along the entire beam path [19]. To correct for this, a background phase image was separately acquired from a water-only region (no cell present) under identical pump/probe conditions, and this background phase signal was subtracted from the photothermal phase images of the cells, following the same background-subtraction procedure described in a previous publication [19]. Representative results for the three mechanisms are shown in Fig. 4b-d. Scattering- and phase-based images showed photothermal contrast across the entire cell body, while fluorescence-based images showed contrast mainly localized to the nucleus. This is evident from the overlay image in Fig. 4e, where the cyan (scattering) and magenta (QPI) signals are visible at the cell membrane and in select intracellular regions outside the nucleus, whereas the yellow (fluorescence) signal is confined largely to the nuclear region, confirming the distinct spatial distributions captured by each contrast mechanism.

### *3.4 Additional data: QPI, scattering, and fluorescence-based mechanisms*

To demonstrate the wide-field MIP capabilities of the technique, a mouse lung tissue section (Fig. 5a) was imaged in scattering mode at 1545 cm⁻¹, corresponding to the protein amide II band. Imaging was performed over a FOV of 216 × 200 µm using a pulse energy of 32 µJ, with images averaged over 2.5 s. A 15 cm focal length (FL) OAP1 was used to increase the FOV. The resulting image reveals the protein distribution within the tissue section (Fig. 5b).

To assess the technique's applicability to smaller biological targets, F-MIP images of E. coli (gram-negative) and S. aureus (gram-positive) were acquired in fluorescence mode, which enables staining and detection of sub-micron structures such as bacteria with strong contrast. Images were acquired using a 7.5 cm FL OAP1 and the DM configuration shown in Fig. 1. Fig. 5c, d show images of SYTO 9-stained *E. coli* and *S. aureus*, respectively, at 1080 cm⁻¹, over a FOV of 103 × 80 µm. Individual *E. coli* and *S. aureus* cells (white rectangles in Fig. 5c, d) are shown magnified in Fig. 5e, f at 1080, 1240, and 1530 cm⁻¹. Images show stronger photothermal contrast at 1530 cm⁻¹ for both species. The photothermal signal ratios (1080/1530 and 1240/1530 cm⁻¹) were calculated for both bacteria, with *S. aureus* showing a higher 1080 cm⁻¹ contrast ratio relative to *E. coli*, as shown in the bar plot in Fig. 5g. Wide-field MIP imaging of transparent SU8 samples with 1.4 µm and 5 µm feature thicknesses was performed using the QPI-based contrast mechanism, with images averaged over 2.5 s. Fig. 5h shows the contrast of the 1.4 µm thick SU8 polymer structures, imaged at 1510 cm⁻¹, a strong absorption band of SU8 [22]. Fig. 5i shows the same target at 5 µm thickness. Different color scales of phase values in Fig. 5h and i indicate the higher phase signal apparent at 5 µm compared to 1.4 µm thickness, demonstrating the QPI channel's ability to distinguish thickness-dependent phase changes, consistent with its quantitative nature. An extended wide-field MIP image of the 1.4 µm thick

SU8 target at 1510 cm⁻¹ is also shown, covering a FOV of 230 × 160 µm (45 µJ pulse energy and 15 cm OAP1 in Fig.1), demonstrating the capability of the technique to image larger areas.

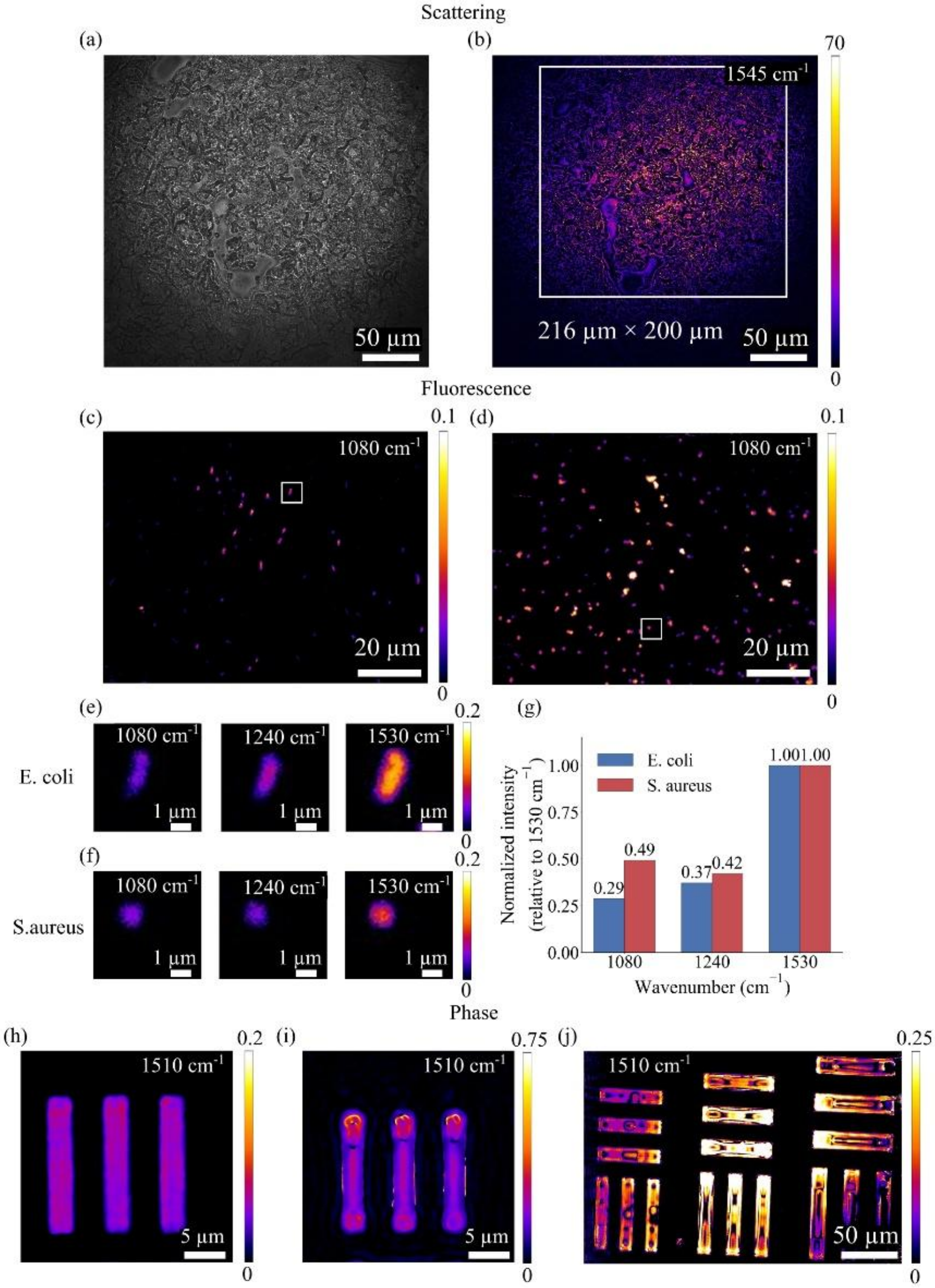

**Fig. 5.** Wide-field multi-contrast imaging of biological and polymer samples. (a, b) Visible and scattering-based photothermal image of a lung tissue section at 1545 cm⁻¹, showing protein distribution across a 216 × 200 µm FOV. (c, d) F-MIP images (at 1 fps, 8 frames averaged) of SYTO 9-stained Gram-negative *E. coli* and Gram-positive *S. aureus* bacteria, respectively, at 1080 cm⁻¹, over a 103 × 80 µm FOV. (e, f) Magnified images of individual *E. coli* and *S. aureus* cells (white rectangles in c, d) at 1080, 1240, and 1530 cm⁻¹. (g) Bar plot of photothermal signal ratios (1080/1530 and 1240/1530 cm⁻¹) for both bacterial species. (h, i) QPI-based photothermal images of SU8 polymer targets with 1.4 µm and 5 µm feature thicknesses, respectively, imaged at 1510 cm⁻¹. (j) Extended wide-field QPI image of the 1.4 µm thick SU8 target at 1510 cm⁻¹, covering a 230 × 160 µm FOV.

## 4. Discussion

In this work, we developed and characterized a wide-field MIP microscope in which photothermal contrast, derived from hot and cold frame acquisition, can be read out through three switchable mechanisms (scattering, fluorescence, and QPI), allowing the readout strategy to be selected according to the sample and imaging task rather than requiring a dedicated instrument for each contrast mechanism. Using PS beads, we showed that all three readout mechanisms show consistent spectral response across the 1410-1510 $cm^{-1}$ range, confirming that spectral information is preserved regardless of the readout mechanism. We further characterized the effective spatial resolution and SNR of each mechanism, and demonstrated multi-contrast imaging of A549 cells in buffer, followed by wide-field protein mapping in mouse lung tissue, F-MIP imaging of *E. coli* and *S. aureus* across multiple wavenumbers, and thickness-dependent phase imaging of SU8 phantom structures, together illustrating the versatility of the multi-contrast platform across both biological and polymer samples.

Table 1 summarizes the measured PSF FWHM, SNR, FOV, and imaging speed of the three contrast mechanisms. Effective spatial resolution, estimated from the measured PSF FWHM using 300 nm PS beads, was comparable between scattering and fluorescence, both below 500 nm; QPI showed more than twice the FWHM. This lower resolution in QPI results from the off-axis geometry used to record the interference pattern. The tilted reference beam introduces a carrier fringe pattern, and separating the true object information from this pattern in the frequency domain requires setting aside part of the camera's usable spatial bandwidth [23]. Because that portion is spent encoding the fringes, less bandwidth remains for capturing fine spatial detail, limiting achievable resolution compared to non-interferometric detection. SNR, estimated from 5 µm PS bead (Fig. 2) images at 1450 $cm^{-1}$, was approximately 26-fold higher for QPI than for scattering. This enhancement is attributed to the interferometric detection scheme, which amplifies the weak photothermally induced phase signal through interference with a strong reference field, providing greater sensitivity to optical path-length changes than direct-intensity detection [24]. Fluorescence showed intermediate SNR, likely reflecting improved background rejection from spectrally distinct emission relative to scattering, while lacking the coherent signal amplification inherent to interferometric detection. Scattering and QPI were detected using a picosecond pulse-width source with a pump-probe delay of less than 100 ns (delay from the length of 20 m MMF), whereas fluorescence was detected with a pulse width below 5 µs. Previous studies have shown that shorter pulse widths and optimized pump-probe delays can improve SNR [8, 13], and differences in acquisition parameters across the three modalities may therefore also contribute to the observed SNR trends. FOV for scattering and QPI was estimated from the wide-field images in Fig. 5b and 5j, acquired using a 50×, 0.75 NA objective and 15 cm OAP1; FOV can be further extended using lower-magnification objectives and alternative OAP1 focal lengths (Fig. 1) to distribute the laser fluence over a larger area. This FOV extension approach, previously demonstrated by distributing higher pulse energy over a wider area to maintain constant fluence, is enabled by the laser's available pulse energy of up to 360 µJ [11, 15, 18, 25], of which less than 50 µJ was used in the present studies. The CMOS camera operated at 20 fps for scattering and QPI, compared to 5 fps or lower for fluorescence, since the lower pulse energy of the LED source requires averaging over more pulses per frame to achieve adequate fluorescence contrast. For the label-free methods, imaging speed could potentially be increased up to 200 fps in future implementations, limited primarily by the 100 Hz repetition rate of the OPO source.

In PS bead imaging (Fig. 2), the bright-center, weak-edge contrast and associated interference fringes in scattering are consistent with the optical behavior of dielectric microspheres under finite-NA imaging: the bead's curved geometry produces a lensing effect that concentrates light toward its center, a pattern also reported in prior wide-field MIP imaging of PS beads [11], and more broadly consistent with nonparaxial Mie-theory models of microsphere image formation under finite NA [26]. In addition, QPI-based images at 1450 cm⁻¹ show artifacts near the bead edge, which may arise from a refractive-index mismatch between the sample and the surrounding air under the dry imaging conditions used here. A similar edge-related artifact is visible in the SU8 target (Fig. 5i, 5 µm thickness) relative to the thinner target (Fig. 5h, 1.4 µm), suggesting the effect becomes more pronounced with increasing sample thickness or larger refractive index-induced phase gradients at boundaries. In contrast, the scattering-based image of the smaller 300 nm bead (Fig. 3) exhibited a single, well-defined central spot, without the bright-center/weak-edge pattern observed for the 5 µm bead. This difference reflects the transition from a wavelength-scale particle, for which the detected field can contain particle-scattering and interference contributions, to a subwavelength particle whose response is approximated by a localized source within the diffraction-limited detection volume [27, 28]. In the latter regime, the measured photothermal contrast is expected to be dominated by the localized thermally induced refractive-index perturbation rather than by spatially resolved boundary-related scattering features.

**Table 1.** Comparison of scattering, fluorescence, and QPI contrast mechanisms in the wide-field MIP microscope.

| Contrast mechanism | Measured PSF FWHM | SNR | FOV and imaging speed | Pros | Cons |
|---|---|---|---|---|---|
| Scattering change | 376 nm | 5 | 216 µm×200 µm (extendable), 20 fps | High spatial resolution, label-free, and simple optical implementation | Lowest SNR of the three mechanisms; prone to lensing-related contrast artifacts in larger samples |
| Fluorescence modulation | 429 nm | 22 | Up to 1 mm [18], 1-5 fps | High specificity, useful for identifying small or specific structures; improved background rejection via spectrally distinct emission | Requires staining/labeling; slower acquisition (<5 fps) |
| Phase change | 904 nm | 128 | 230 µm × 160 µm (extendable), 20 fps | Highest SNR; label-free measurement of optical path length/thickness | Lower spatial resolution; edge-related artifacts observed near sample boundaries, in thicker samples under dry conditions |

Besides the SU8 polymer and PS beads, we imaged A549 cells stained with SYTO 9 in buffer using all three contrast mechanisms (Fig. 4). Scattering-based imaging showed contrast both at the cell border and within the intracellular region. The border contrast is consistent with mid-IR absorption of the $PO_2^-$ head groups of membrane phospholipids at 1240 cm⁻¹ [29, 30], enhanced by the strong refractive-index gradient at the cell boundary, while the intracellular contrast likely reflects other phosphate-containing constituents, such as cytoplasmic RNA and organelle membranes, distributed throughout the cell body [31, 32]. The same phospholipid contribution accounts for the membrane signal seen in the QPI channel (Fig. 4d) [33]. Fluorescence imaging, acquired at 1240 $cm^{-1}$, showed contrast only from the nucleus (Fig. 4c), as expected, since SYTO 9 labels nucleic acids and F-MIP reports IR absorption only in the vicinity of the fluorophore. QPI imaging showed contrast from both the nuclear region and the cell border; however, portions of the border were not consistently visible across the image, and the underlying cause of this incomplete border contrast is not yet clear. We also observed a weaker water/buffer background in the scattering and fluorescence channels than in QPI. This is likely because QPI measures the optical phase accumulated along the optical path through the sample and is therefore sensitive to spatially extended refractive-index changes produced by thermo-optic and thermal-expansion effects in the heated aqueous medium [19, 34]. By contrast, scattering detection is primarily sensitive to localized refractive-index heterogeneities and interfaces; a nearly uniform refractive-index change in the surrounding buffer produces negligible additional scattering, although residual contributions from interfaces, gradients, and background scatterers may remain [28, 35]. Fluorescence detection is also comparatively insensitive to the nonfluorescent water/buffer matrix [36].

Additionally, *E. coli* and *S. aureus* were imaged using F-MIP imaging. The photothermal signal ratio (1080/1530 $cm^{-1}$), calculated from the average pixel intensity within the bacterial area, was higher for *S. aureus* than for *E. coli* (Fig. 5g). This difference is consistent with cell wall composition between the Gram-positive and Gram-negative species, which is well known from FTIR and prior AFM-IR studies showing that spectral bands originating from cell wall components such as peptidoglycan and teichoic acid distinguish Gram-positive from Gram-negative bacteria [37]. Establishing robust species discrimination requires to collect hyperspectral data at sufficient small wavenumber intervals followed by reconstruction of spectra and statistical analysis of replicate measurements across larger sample sets.

A few limitations of the current setup should be noted. First, the fluorescence pathway uses a dedicated DM for most samples, but this requires manual switching from the shared BS used by the other channels, rather than an automated transition between imaging modes. The 5 μm PS bead was the only sample imaged with the shared BS instead of the DM, feasible only due to its strong fluorescence; weaker-fluorescing samples require the dedicated DM for reliable collection. Replacing manual switching with an automated mechanism would improve collection efficiency and eliminate manual intervention. Second, the OPO's internal IR attenuation is coupled to the 1064 nm pump pulse energy, so adjusting IR attenuation also alters the pump energy delivered to the visible detection channels; a separate visible probe source, decoupled from the OPO pump, may resolve this while improving control over the pump-probe delay. A higher technological readiness level is targeted to implement this multi-contrast MIP microscope platform for screening bacteria, cells, and tissue sections as an attractive tool for clinical application.

## 5. Conclusion

We presented a wide-field MIP microscope in which scattering, fluorescence, and quantitative phase readout can be switched within a single platform, allowing the detection strategy to match the sample and task at hand rather than being fixed to one contrast mechanism. By validating the platform on polymer and biological samples, we demonstrate chemically specific imaging across diverse material and biological contexts, combining label-free scattering- and phase-based detection with F-MIP imaging. This flexibility positions the instrument for future studies requiring adaptable contrast, with ongoing work aimed toward simultaneous multi-contrast acquisition.

**Funding.** BMFTR (Federal Ministry of Research, Technology and Space) funding program Photonics Research Germany (13N15464), integrated into the Leibniz Center for Photonics in Infection Research (LPI).

**Acknowledgments:** The authors thank Thomas Cismar and Andre Gomez for providing the 300 nm PS beads, Uwe Hübner for preparing the SU8 resolution targets (all from Leibniz IPHT), Susanne Pahlow for helping with bacterial preparation (Leibniz IPHT), and Artem Shydliukh (Leibniz IPHT) and Stefanie Deinhardt-Emmer (University Hospital Jena) for providing the lung tissue sections.

**Ethics statement:** The mice experiments, from which a lung tissue section was obtained, were approved by the local ethics committee of the Thuringian State Office for Consumer Protection (TV-Nr: 02-018/16; twz05-2022).

**Disclosures.** The authors declare no competing interests.

**Data availability.** The data underlying the results presented in this paper are available from the authors upon reasonable request (e-mail: christoph.krafft@leibniz-ipht.de).